\documentclass[]{zakoproc}
\usepackage{graphicx}
\begin{document}

\title{Magnetospheric launching in resistive MHD simulations}
\author{Miljenko \v{C}emelji\'{c} and Hsien Shang}
\institute{Academia Sinica, Institute of Astronomy and Astrophysics and
Theoretical Institute for Advanced Research in
Astrophysics, P.O. Box 23-141, Taipei 106, Taiwan }
\markboth{M. \v{C}emelji\'{c}, H. Shang}{Magnetospheric launching \ldots}

\maketitle

\begin{abstract}
We perform numerical simulations in the close vicinity of a slowly
rotating young stellar object. Using our own resistive MHD Zeus347 code in
2D axisymmetry, magnetospheric interaction experiences four robust stages
in evolution. Quasi-stationary axial and conical streams of outflowing
matter can last many orbital periods as results of the
resistivity-facilitated magnetic reconnections. The shape of the magnetic
field depends on resistivity in the magnetosphere.

\end{abstract}

\section{Setup and Results}
\begin{figure}[ht]
 \includegraphics[width=.35\textwidth]{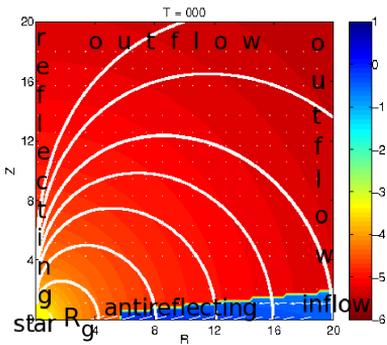}
 \caption{In logarithmic color grading is shown the density in our initial
 conditions, solid lines show the poloidal magnetic field of a stellar
dipole, and vectors depict the velocity. Star is set as a rotating,
absorbing boundary condition inside the computational box, enclosing the
origin. In the disk gap, matter is allowed to flow through the disk
mid-plane.
}
\label{fig1}
\end{figure}

\begin{figure}[ht]
 \includegraphics[width=.43\textwidth]{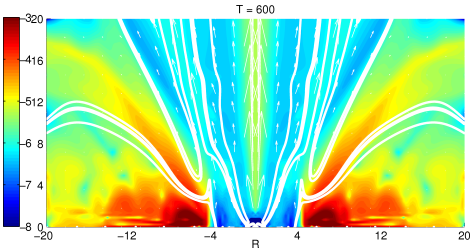}
 \caption{
Final stage in our simulations with stellar magnetic field of the order of
100 G. In logarithmic color grading is shown the poloidal mass flux, the
solid lines show poloidal magnetic field lines, and arrows show poloidal
velocity vectors. Conical and axial outflows are ejected from the close
vicinity of the star.
}
\label{fig2}
\end{figure}

In simulations with our resistive version of the Zeus-3D code, in the
axisymmetry option, we set innermost region of a star-disk system
(see Fig.~\ref{fig1}). The computational domain is
$R\times Z=(90\times 90)\mathrm{~grid\ cells}=(0.2\times 0.2)$~AU. Disk is
in a slightly sub-Keplerian rotation, with a rotating, hydrostatic corona
and purely dipole magnetic field centered at the origin. Resistivity in
the disk is constant, and in corona it is modeled by density, with
$\eta\sim\rho^{1/3}$, following \cite{FC02}.
All simulations of star-disk interaction in our setup go through four
stages: 1) relaxation with pinching of magnetic field inwards, 2)
reconnection with opening of the stellar dipole, 3) narrowing of the disk
gap with formation of transient funnel flow onto the stellar surface, 4)
final stage of equilibrium of magnetic and disk ram pressure, with two
outflows, one axial and another conical. The final stage is shown in
Fig.~\ref{fig2}. It is similar to the result reported in \cite{R09}, but
with the difference that our simulation is in the regime with magnetic
Prandtl number smaller than unity. Physical resistivity, together with
reconnection, helps launching of outflows.
\begin{figure}[ht]
 \includegraphics[width=.55\textwidth]{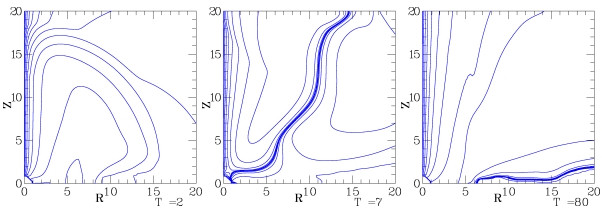}
 \caption{Poloidal magnetic field lines for the different timesteps in the
same simulation. Reshaping of the initial dipolar magnetic field is
enabled by reconnection. After the pinching of the field in plasmoid
ejected during the relaxation ({\it left} panel), reconnection helps the
change of field topology ({\it middle} panel) into the stellar and disk
field components ({\it right} panel).
}\label{figure3}
\end{figure}
\section{Conclusions}
In purely resistive MHD numerical simulations, we obtain solutions with
two streams of matter flowing out from the innermost magnetosphere around a
young stellar object. This is the first simulation in which a conical
outflow is launched with magnetic Prandtl number less than unity, that is,
with resistivity larger than viscosity, by magnetic reconnection. The
configurations of the field are modified during the reconnection to
channel flowing out from the innermost region of the disks. 

\end{document}